\documentclass[
reprint,
superscriptaddress,
amsmath,
amssymb,
aps,
longbibliography,
pra,
showpacs,
floatfix
]{revtex4-1}
\usepackage[colorlinks=true,linkcolor=blue,citecolor=blue,urlcolor=blue]{hyperref}
\usepackage{graphicx}
\usepackage{xcolor} 

\usepackage{tabularx}
\usepackage{multirow}

\usepackage{braket}

\usepackage{comment}

\usepackage{pifont}
\newcommand{\cmark}{\ding{51}}%
\newcommand{\xmark}{\ding{55}}%

\begin{document}

\title{Phonon-induced geometric chirality}

\author{Carl~P.\ Romao}
\email{carl.romao@mat.ethz.ch}
\affiliation{Department of Materials, ETH Zurich, CH-8093 Zurich, Switzerland}

\author{Dominik~M.\ Juraschek}
\email{djuraschek@tauex.tau.ac.il}
\affiliation{School of Physics and Astronomy, Tel Aviv University, Tel Aviv 6997801, Israel}

\date{\today}


\begin{abstract}
Chiral properties have seen increasing use in recent years, leading to the emerging fields of chiral quantum optics, plasmonics, and phononics. While these fields have achieved manipulation of the chirality of light and lattice vibrations, controlling the chirality of materials on demand has yet remained elusive. Here, we demonstrate that linearly polarized phonons can be used to induce geometric chirality in achiral crystals when excited with an ultrashort laser pulse. We show that nonlinear phonon coupling quasistatically displaces the crystal structure along phonon modes that reduce the symmetry of the lattice to that of a chiral point group corresponding to a chiral crystal. By reorienting the polarization of the laser pulse, the two enantiomers can be induced selectively. Therefore, \textit{geometric chiral phonons} enable the light-induced creation of chiral crystal structures and offer a pathway to engineering chiral electronic states and optical properties.
\end{abstract}

\maketitle


\section*{Introduction}

Chirality describes a fundamental asymmetry of geometric objects and plays a central role in chemistry, biology, and physics.  A geometric object is chiral when its mirror image cannot be superposed on the original system by any combination of rotations and translations. The chiral geometric arrangement of atoms in molecules and solids determines chemical reactivity and electronic phases, and is one of the central requirements for the formation of life \cite{Bonner1995,Barron2008}. Geometric, or static, chirality can be extended to time-dependent systems, where a sense of handedness is applied to moving and rotating objects \cite{Barron1986}. The interplay of geometric and dynamic chirality can, for example, be observed in the chirality-induced spin selectivity (CISS) effect, which has been hypothesized to have influenced the origin of biological geometric homochirality by inducing enantioselective crystallization of RNA \cite{ozturk2023origin}, and has recently been connected to chiral phonons \cite{ohe2024chirality}. Chiral phonons represent such a case of motion-based, or dynamic, chirality, as they conventionally describe circularly polarized lattice vibrations that carry linear and angular momentum and can therefore be assigned a well-defined helicity \cite{zhang:2015,Zhu2018,Ishito2023,Ueda2023}. This angular momentum of chiral phonons has in recent years been shown to lead to a plethora of emergent phenomena, including phonon Hall \cite{Grissonnanche2020,Chen2022_phononHall,Flebus2022}, phonon Zeeman \cite{juraschek2:2017,Juraschek2019,Cheng2020,Baydin2022}, and phonon Einstein-de Haas effects \cite{zhang:2014,Dornes2019,Tauchert2022}, as well as equilibrium and ultrafast  phono-magnetism \cite{nova:2017,juraschek2:2017,Juraschek2019,Juraschek2020_3,Geilhufe2021,Basini2024,Davies2024,Luo2023,Kahana2023,Hamada2020,Ren2021,Zabalo2022,Chaudhary2023,Wu2023,Zhang2023bilayergraphene,Cui2023}. An intriguing question therefore arises of whether not just dynamic, but geometric chirality can be induced with lattice vibrations, which would create a path to dynamically inducing electronic and optical properties that are connected to the chiral structure of materials \cite{Hentschel2017,Lodahl2017,Meskers2022}.

\begin{figure*}[t]
\centering
\includegraphics[scale=0.100]{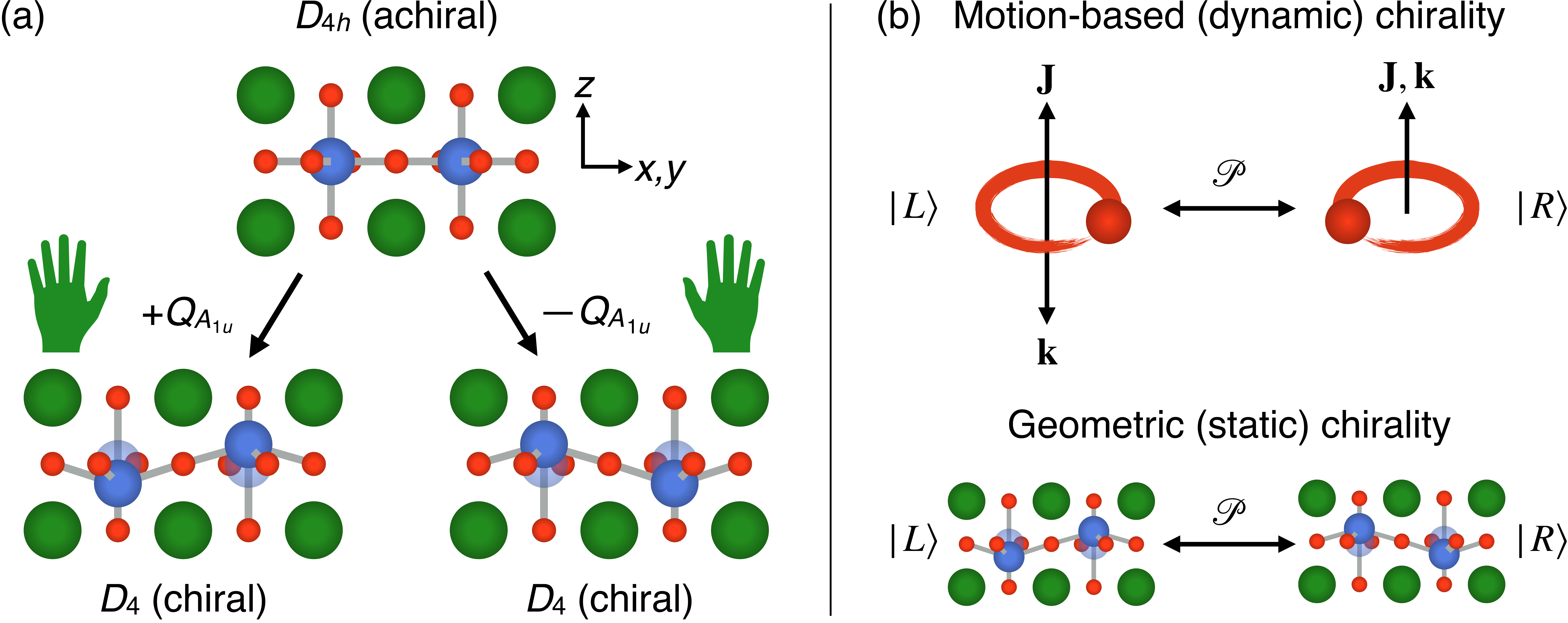}
\caption{
\label{fig:concept}
\textbf{Phonon-induced geometric chirality.} (a) We show the example of a tetragonal $AB$O$_3$ perovskite in the achiral point group $D_{4h}$. The $A_{1u}$ mode moves the $B$-site cations along the tetragonal axis ($z$) and breaks all improper rotation symmetries, including inversion and mirror, which reduces the symmetry of the system to the chiral point group $D_4$. Positive and negative displacements, $\pm Q_{A_{1u}}$, lead to two enantiomers. (b) In conventional chiral phonons, the circular motion of the atoms produces an angular momentum, $\mathbf{J}$, which combined with the propagation direction of the phonon, $\mathbf{k}$, breaks all improper rotation symmetries \cite{Ishito2023,Ueda2023}. This is called motion-based (or dynamic) chirality \cite{Barron1986}. In geometric chiral phonons, as discussed in this study, the displacement of the atoms itself, not the motion, breaks all improper rotation symmetries, leading to geometric (or static) chirality. In both cases, the two enantiomeric states, $\vert L \rangle$ and $\vert R \rangle$, are connected by the parity operation, $\mathcal{P}$.}
\end{figure*}

Here, we demonstrate theoretically and computationally that achiral crystals can be made chiral through the nonlinear excitation of phonon modes by intense ultrashort laser pulses. We perform a group-theoretical analysis to identify phonon modes whose displacement pattern breaks all improper rotation symmetries, including inversion and mirror, and therefore leads to a chiral atomic structure (Fig.~\ref{fig:concept}a). We show that the crystal lattice can be quasistatically distorted along the eigenvectors of these phonon modes through the mechanism of nonlinear phononic rectification, which offsets the average of the atomic vibrations from their respective equilibrium positions, $\langle Q_c\rangle\neq 0$. The direction of displacement can be controlled through the polarization of the laser pulse, making chirality switching between the two enantiomers possible. We quantify the degree of induced chirality by introducing an electro-chiral susceptibility and we evaluate the chiral response to the excitation by an ultrashort pulse for the example of lithium triborate (LiB$_3$O$_5$).


\section*{Results}


\subsection*{Geometric chiral phonons}

We begin with a group-theoretical analysis of phonon modes that are able to induce chiral crystal geometries. Chiral crystals do not contain any axis of improper rotation, including inversion, mirrors, and rotoreflections. They are characterized by the 65 Sohncke space groups, which in turn are part of the 11 chiral point groups, shown in Supporting Table~\ref{tab:sohncke} \cite{Fecher2022}. In achiral crystals, there may exist phonon modes whose displacement pattern reduces the symmetry of the lattice to one of the Sohncke groups, therefore inducing a chiral crystal geometry. These phonons are to be contrasted with the circularly polarized chiral phonons commonly found in literature \cite{Ishito2023,Ueda2023}, where the motion of the atoms, not the displacement itself, leads to chirality, as illustrated in Fig.~\ref{fig:concept}b. We will therefore refer to these phonon modes as \textit{geometric chiral phonons} in the remainder of the manuscript. We will focus on phonon modes at the center of the Brillouin zone (long-wavelength limit), which can be excited with light. In our analysis, we utilize the character and multiplication tables of the Bilbao Crystallographic Server \cite{Bilbao2006}, and extract the group--subgroup relations corresponding to the phonon displacements using the ISOSUBGROUP software tool \cite{Stokes2016}. 

In Supporting Tables~\ref{tab:phonons1a} and \ref{tab:phonons1b}, we show a list of the 32 point groups and indicate whether they possess inversion ($\mathcal{I}$) or mirror ($\mathcal{M}$) symmetries. The 11 chiral point groups possess neither inversion nor mirror symmetries. Point group $S_4$($\bar{4}$) is a special case, which does not contain an inversion center or mirror planes, but contains a rotoreflection symmetry and is therefore not chiral. In the tables, we list the irreducible representations of phonon modes that break all inversion and mirror symmetries, as well as the respective subgroup symmetries they induce. Nearly all of them lead to chiral point groups and can therefore be considered geometric chiral phonons. Exceptions are the $B_u$ modes of the $C_{4h}$(4/m) point group, as well as the $A_2$ modes of the $D_{2d}$($\bar{4}$2m) point group, which break the inversion symmetry and all mirror symmetries, but which reduce the crystal symmetry to $S_4$($\bar{4}$). For doubly and triply degenerate phonon modes, the induced subgroup depends on the chosen basis of the eigenvectors and we list all possible outcomes in the tables.




\subsection*{Time-averaged geometric chirality}

We will next examine how geometric chiral phonons can be excited by an ultrashort laser pulse to yield a macroscopic chiral crystal geometry. Harmonic excitations of geometric chiral phonons would leave the time-averaged crystal structure in its achiral state, where the atoms vibrate uniformly around their respective equilibrium positions. We therefore utilize nonlinear phononic rectification \cite{forst:2011,subedi:2014} as a means to unidirectionally distort the lattice along the eigenvectors of a geometric chiral phonon mode. This can be achieved through three-phonon coupling of the type $Q_a Q_b Q_c$, where $Q_a$ and $Q_b$ are the amplitudes of phonon modes that are coherently driven by an ultrashort laser pulse and that couple nonlinearly to a geometric chiral phonon, $Q_c$. The mean amplitude of the geometric chiral phonon then follows the mean-square amplitude of the driven phonons, $\langle Q_c\rangle \propto \langle Q_a Q_b\rangle \neq 0$, which leads to a quasistatic shift of the geometric chiral phonon, as long as the frequencies of the $a$ and $b$ modes are reasonably close to each other. As a result, a chiral crystal geometry can be induced. Positive and negative displacements along $Q_c$ are related by mirror symmetry and therefore correspond to two enantiomers.

In Supporting Tables~\ref{tab:phonons1a} and \ref{tab:phonons1b}, we list all symmetry-allowed three-phonon couplings that contain the geometric chiral phonon. In order for a coupling to be symmetry-allowed, it needs to contain the fully symmetric representation. For nondegenerate phonon modes, $Q_a$ and $Q_b$ can correspond to the same ($a=b$) or to different irreducible representations ($a\neq b$). For doubly and triply degenerate phonon modes, they generally correspond to the two orthogonal components of the same irreducible representation, e.g. $Q_a\equiv Q_{a(1)}$ and $Q_b\equiv Q_{a(2)}$. In order to achieve rectification with nondegenerate phonon modes, the eigenfrequencies of the two modes, $\Omega_a$ and $\Omega_b$, should be close together, so that the dephasing time of the two components is longer than the respective phonon lifetimes, $\kappa_a$ and $\kappa_b$, $2\pi(\Omega_a-\Omega_b)^{-1} > \kappa_{a/b}^{-1}$.

Achieving a measurable rectification of the crystal structure, $\langle Q_c\rangle \neq 0$, relies on excitation of the driven phonon modes, $Q_a$ and $Q_b$, with large amplitudes, which can generally be achieved with resonant IR absorption, but not with impulsive stimulated Raman scattering. We therefore list the irreducible representations corresponding to the IR-active phonon modes in Supporting Tables~\ref{tab:phonons1a} and \ref{tab:phonons1b}. In centrosymmetric materials, geometric chiral phonons break inversion symmetry and can only couple to one infrared (IR)- and one Raman-active phonon mode at a time. In noncentrosymmetric materials, inversion symmetry is already broken, which lifts the principle of mutual exclusion for IR and Raman activity and allows geometric chiral phonons to couple to two IR-active phonon modes. Therefore, we expect the effect of phonon-induced geometric chirality to be most easily achievable in noncentrosymmetric materials and we will focus our analysis on these cases. 

In Table~\ref{tab:phonons1c}, we show the 10 noncentrosymmetric-achiral point groups, in which three-phonon couplings of geometric chiral phonons to IR-active modes are possible. We further list example materials for each of the point groups in which these couplings can be found. We have limited our analysis to group--subgroup relations of point groups, as the emergence of chirality depends only on the point group. The corresponding relations for space groups can be determined using the ISODISTORT and ISOSUBGROUP software packages where necessary \cite{Stokes2016}. An analysis of the group--subgroup relations of space groups yields that the displacement induced by geometric chiral phonons at the Brillouin-zone center always leads to one of the 43 nonenantiomorphic Sohncke groups and not to one of the 11 enantiomorphic space-group pairs. Pairs of enantiomers in the nonenantiomorphic Sohncke groups have no unambiguous ``left'' or ``right'' handedness defined by their symmetry operations. These pairs therefore represent a case of nonhanded chirality \cite{king2001nonhanded, king2003chirality}.


\begin{table*}[t]
\centering
\begin{tabular}{llllll} 
\hline\hline
Point group~  & Chiral mode~ & Subgroup~ & IR-active modes~ & Nonlinear coupling & Example materials \\
\hline\hline
$C_s$(m)  & $A''$  & $C_1$(1) & $A'(x,y)$, $A''(z)$ & $A'' A'' A'$ & KH$_2$PO$_4$, BaGa$_4$Se$_7$ \\ 
$C_{2v}$(mm2)  & $A_2$  & $C_2$(2) &  $B_1(x)$, $B_2(y)$ & $A_2 B_1 B_2$ & LiB$_3$O$_5$, CdTiO$_3$, GaFeO$_3$ \\
$S_4$($\bar{4}$) &  $B$ &  $C_2$(2) & ${}^{i}E(x,y)$ &  $B{}^{i}E{}^{i}E$  ($i=1,2$) & BPO$_4$, Na$_2$ZnSnS$_4$  \\
 & ${}^{i}E$ &  $C_1$(1)  & ${}^{i}E(x,y)$, $B(z)$  &   ${}^{i}E{}^{i}EB$ ($i=1,2$) &    \\
$C_{4v}$(4mm)  & $A_2$  & $C_4$(4)  & $E(x,y)$ & $A_2 E E$ & SrBaNb$_2$O$_6$ \\ 
 & $E$ & $C_1$(1), $C_s$(m)$^*$ & $E(x,y)$, $A_1(z)$ & $E E A_1$ &  \\
$D_{2d}$($\bar{4}$2m)  & $A_2$ & $S_4$($\bar{4}$)$^*$ & $E(x,y)$ & $A_2 E E$ & ZnGeP$_2$, AgGaS$_2$  \\
 & $B_1$ & $D_2$(222) & $E(x,y)$ & $B_1 E E$ &   \\
 & $E$ & $C_1$(1), $C_2$(2), $C_s$(m)$^*$ & $E(x,y)$, $B_2(z)$ & $E E B_2$ &  \\
$C_{3v}$(3m)  & $A_2$ & $C_3$(3) & $E(x,y)$  & $A_2 E E$ & LiNbO$_3$, BiFeO$_3$ \\ 
  & $E$ & $C_1$(1), $C_s$(m)$^*$ & $E(x,y)$, $A_1(z)$ & $E E A_1$, $E E E$ &  \\
$C_{3h}$($\bar{6}$)  & ${}^{i}E''$ &  $C_1$(1) & ${}^{i}E'(x,y)$, $A''(z)$ & ${}^{i}E'' {}^{j}E' A''$ ($i,j=1,2$) & LiCdBO$_3$, BaZnBO$_3$F  \\
$C_{6v}$(6mm)  & $A_2$ & $C_6$(6) & $E_1(x,y)$ & $A_2 E_1 E_1$ & GaBO$_3$, Ba$_3$YbB$_3$O$_9$  \\
 & $E_1$ & $C_1$(1), $C_s$(m)$^*$ & $E_1(x,y)$, $A_1(z)$ & $E_1 E_1 A_1$ &  \\
 & $E_2$ & $C_2$(2), $C_{2v}$(mm2)$^*$ & $E_1(x,y)$ & $E_2 E_1 E_1$ &  \\
$D_{3h}$($\bar{6}$m2)  & $E''$ & $C_1$(1), $C_2$(2), $C_s$(m)$^*$ & $E'(x,y)$, $A_2''(z)$  & $E'' E' A_2''$  & Na$_3$La$_9$B$_8$O$_{27}$  \\ 
$T_d$($\bar{4}$3m)  & $E$ & $D_2$(222), $D_{2d}$($\bar{4}$2m)$^*$  & $T_2(x,y,z)$ & $E T_2 T_2$ & CsNbMoO$_6$, Zn$_4$B$_6$O$_{13}$  \\ 
& $T_1$ & $P_1$(1), $C_s$(m),$^*$ $S_4$($\bar{4}$),$^*$ $C_3$(3) & $T_2(x,y,z)$ & $T_1 T_2 T_2$ &  \\
\multicolumn{6}{l}{{}$^{*}$Subgroup is not chiral} \\
\hline\hline
\end{tabular}
\caption{\textbf{Geometric chiral phonons in noncentrosymmetric-achiral point groups.} We show the 10 noncentrosymmetric-achiral point groups, in which geometric chiral phonons can couple to two IR-active phonon modes, which makes their excitation through nonlinear phononic rectification feasible. For each point group, we list the irreducible representations of the geometric chiral phonons, the chiral subgroups that their displacement leads to, as well as the IR-active phonon modes they couple to and their coupling terms. We further show example materials for each of the cases, specifically, 
 example materials which can be produced as optical single crystals. For doubly and triply degenerate modes, the induced subgroup depends on the chosen basis of eigenvectors, which may also lead to achiral subgroups.}
\label{tab:phonons1c}
\end{table*}


\subsection*{Nonlinear phonon dynamics}

The coherent phonon dynamics following excitation by an ultrashort terahertz or mid-IR pulse can be captured in a semi-classical oscillator model, for which the input parameters can be computed from first principles \cite{subedi:2014,fechner:2016,juraschek:2017}. This has in the past shown to yield quantitatively accurate estimates of the phonon amplitudes for semiconductors and insulators, where the photon energy of the laser pulse is far below the band gap \cite{VonHoegen2018,Henstridge2022}. A minimal model of three-phonon coupling that captures nonlinear phononic rectification can be written as
\begin{equation}\label{eq:minimalpotential}
V_{\textrm{ph}} = \frac{\Omega^2_a}{2}Q_a^2 + \frac{\Omega^2_b}{2}Q_b^2 + \frac{\Omega^2_c}{2}Q_c^2 + g Q_a Q_b Q_c,
\end{equation}
where $\Omega_\alpha$ is the eigenfrequency of phonon mode $\alpha$, and where $g$ is the nonlinear phonon coupling coefficient. The phonon amplitudes, $Q_\alpha$, are given in units of \AA$\sqrt{u}$, where $u$ is the atomic mass unit. The light-matter coupling of the laser pulse to the IR-active phonons can be described as $V_{\text{l-m}} = -\mathbf{p}_\alpha\cdot\mathbf{E}(t) = - Z_{\alpha,i} Q_\alpha E_i(t)$, where $\mathbf{p}_\alpha=\mathbf{Z}_\alpha Q_\alpha$ is the electric dipole moment of phonon mode $\alpha$ and $\mathbf{E}$(t) is the electric field component of the laser pulse. $\mathbf{Z}_\alpha = \sum_n Z^*_n \mathbf{q}_{\alpha,n}/\sqrt{M_n}$ is the mode effective charge vector, where $Z^*_n$ is the Born effective charge tensor of atom $n$, $\mathbf{q}_{\alpha,n}$ is its unitless phonon eigenvector, and $M_n$ is its atomic mass. The electric field component of the laser pulse can be expressed as $\mathbf{E}(t)=(E(t) \sin(\phi),E(t) \cos(\phi),0)$, where $\phi$ is an angle defining the orientation of linear polarization. $E(t) =  E_0 \exp[-t^2/(\tau \sqrt{8 \mathrm{ln (2)}})^2] \cos (\omega_0 t)$, where $E_0$ is the peak electric field, $\tau$ is the full width at half maximum pulse duration, and $\omega_0$ is the center frequency. We use the Einstein sum convention for the spatial index $i\in\{x,y,z\}$. The equations of motion can be written as 
\begin{equation}
    \ddot{Q}_\alpha + \kappa_\alpha \dot{Q}_\alpha + \partial_{Q_\alpha}V  = 0, \label{eq:phononeom}
\end{equation}
where $\kappa_\alpha$ are the phonon linewidths and $V=V_{\text{ph}}+V_{\text{l-m}}$. The solutions of Eqs.~\eqref{eq:phononeom} can be approximated to first and second order in the electric field for the IR-active and geometric chiral phonon modes, respectively \cite{Kahana2023}, see Methods.


\subsection*{Electro-chiral susceptibility}

Chirality can be quantified through various distance-based measures that compare an achiral parent structure with a corresponding symmetry-broken chiral structure. Here, we apply the continuous chirality measure \cite{Zabrodsky1995,Fecher2022}, which is defined as
\begin{equation}\label{eq:chiralitymeasure}
s = \sqrt{\sum\limits_n ||\mathbf{r}_n-\mathbf{r}_{n,\mathrm{achiral}}||^2}.
\end{equation}
where $\mathbf{r}_{n,\mathrm{achiral}}$ is the position of atom $n$ in the unit cell of the achiral parent structure and $\mathbf{r}_n$ is a continuous position variable between the achiral and chiral structures. The index $n$ runs over all atoms in the unit cell. The continuous chirality measure can be normalized as $S = s/N$ with $N = (\sum_n ||\mathbf{r}_{n,\mathrm{chiral}}-\mathbf{r}_{n,\mathrm{achiral}}||^2)^{1/2}$, where $\mathbf{r}_{n,\mathrm{chiral}}$ is the position of atom $n$ in the chiral structure. 
In our analysis, $\mathbf{r}_{n,\mathrm{achiral}}$ corresponds to the positions of the atoms in the equilibrium structure, whereas $\mathbf{r}_n$ corresponds to the time-dependent atomic coordinates along the eigenvectors of the geometric chiral phonon mode. The norm of the atomic displacements from the equilibrium structure is then simply given by $||\mathbf{r}_n-\mathbf{r}_{n,\mathrm{achiral}}||=  ||Q_c \mathbf{q}_{c,n}||/\sqrt{M_n}$. Generally, one cannot assume that an achiral material also has a corresponding stable chiral phase (or vice-versa), which makes $\mathbf{r}_{n,\mathrm{chiral}}$ undefined. In order to still define a measure, we select $\mathbf{r}_{n,\mathrm{chiral}}$ as the positions of the atoms corresponding to the displace from the equilibrium structure with unit eigenvectors, $||\mathbf{r}_{n,\mathrm{chiral}}-\mathbf{r}_{n,\mathrm{achiral}}||= ||\mathbf{q}_{c,n}||/\sqrt{M_n}$. With these considerations, we can express a time-dependent and mode-resolved normalized continuous chirality measure that is simply proportional to the unitless amplitude norm of the geometric chiral phonon, $S_c(t) = ||Q_c(t)||/(\text{\AA}\sqrt{u})$. The conventional definition of the continuous chirality measure in Eq.~\eqref{eq:chiralitymeasure} only returns positive values and therefore doesn't capture transitions between enantiomers. Positive and negative displacements along $Q_c$ correspond to two enantiomers, however, and we need a way to distinguish them. We therefore drop the norm in our following definition of the continuous chirality measure, which then simply reads
\begin{equation}\label{eq:chiralitymeasurephonon}
    S_c(t) = \frac{1}{\text{\AA}\sqrt{u}}Q_c(t).
\end{equation}
We can now insert the solutions for the phonon amplitudes from Eqs.~\eqref{eq:phononeom} (see Methods) into the Fourier transform of Eq.~\eqref{eq:chiralitymeasurephonon}, which yields a frequency-dependent continuous chirality measure,
\begin{align}
S_c(\omega) & = \int_\infty^\infty \chi^{(2)}_{c,ij}(\omega,\omega')E_i(\omega-\omega')E_j(\omega')\mathrm{d}\omega', \label{eq:chiralitymeasureintegral} \\
 \chi^{(2)}_{c,ij}(\omega,\omega') & = \frac{1}{\text{\AA}\sqrt{u}}\frac{-gZ_{a,i}Z_{b,j}}{\Delta_c(\omega)\Delta_a(\omega-\omega')\Delta_
b(\omega')}, \label{eq:electrochiralsusceptibility}
\end{align}
where $\Delta_{\alpha}(\omega)=\Omega_{\alpha}^2 - \omega^2 + i\omega\kappa_{\alpha}$. $\chi^{(2)}_{c,ij}(\omega,\omega')$ can be seen as a mode-resolved electro-chiral susceptibility that is a measure for the frequency-resolved response of chiral displacements to the excitation by the laser pulse. The zero-frequency component of the susceptibility, $\chi^{(2)}_{c,ij}(\omega=0,\omega')$ can be used as an estimate of whether nonlinear phononic rectification and therefore phonon-induced chirality is achieved, without having to solve the equations of motion. For resonant pumping of the IR-active phonon modes, $\omega'\approx\Omega_a , \Omega_b$. If multiple geometric chiral phonon modes are involved in the dynamics, Eq.~\eqref{eq:chiralitymeasure} contains a sum over them, see Methods.


\begin{figure*}[t]
\centering
\includegraphics[width = 0.9\textwidth,trim=1.5cm 5cm 3cm 1cm]{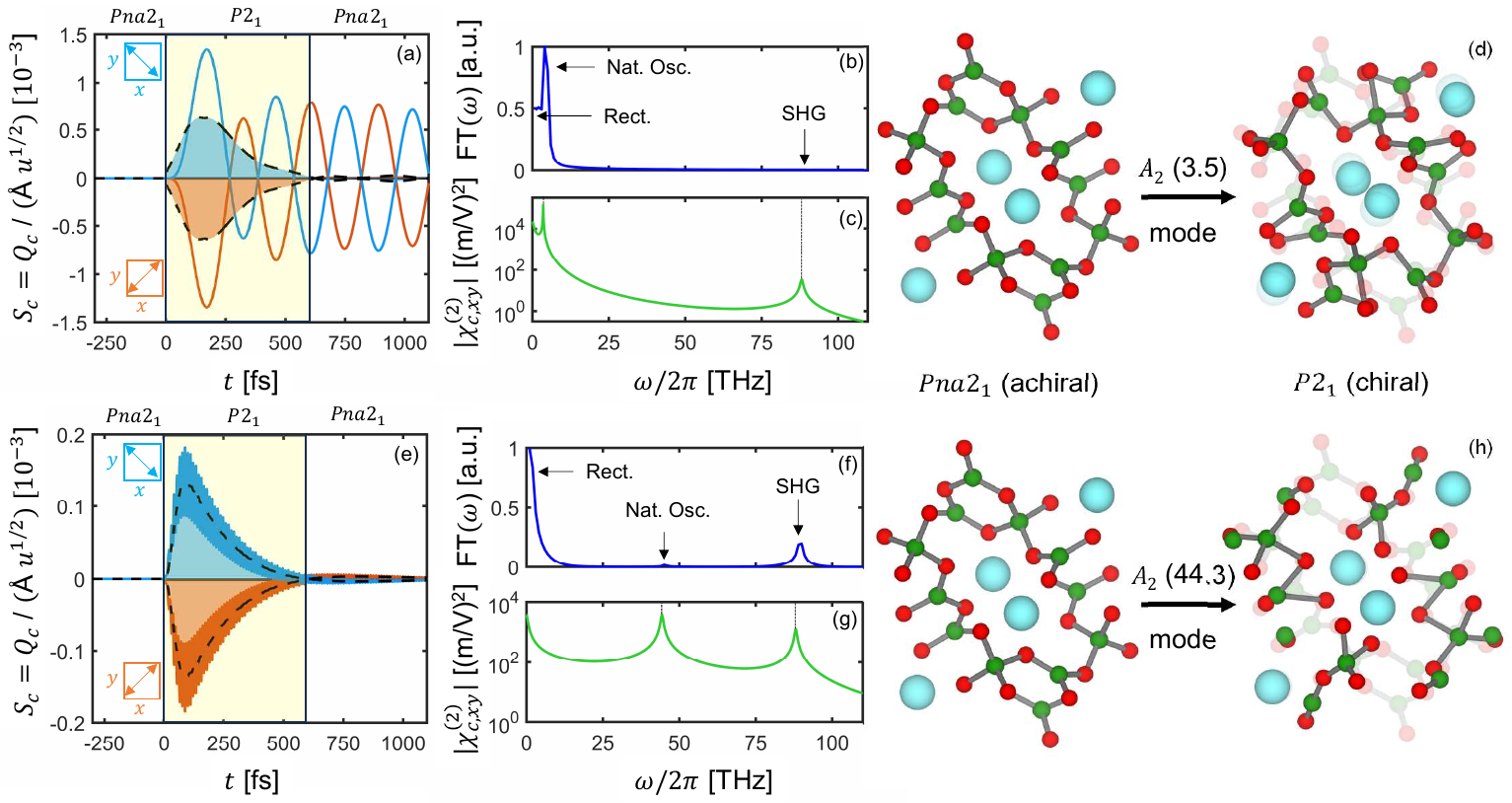}
\caption{
\label{fig:examples}
\textbf{Phonon-induced geometric chirality in LiB$_3$O$_5$.} Nonlinear excitation of the geometric chiral $A_2$ modes at 3.5 (a-d) and 44.3~THz (e-h). 
(a,e) Time evolution of the $A_2$ phonon amplitudes, $Q_c$, (blue and orange lines) and mean continuous chirality measures, $\langle S_c \rangle$, (black dotted lines) following the resonant excitation of the IR-active $B_1$(44.5) and $B_2$(44.0) modes by an ultrashort mid-IR pulse. 
Reorientation of the polarization of the laser pulse by 90$^\circ$ with respect to the crystal axes (insets) switches the sign of $Q_c$ and $\langle S_c \rangle$, making it possible selectively create the two enantiomers. 
(b,f) Fourier transforms of the phonon amplitudes, revealing quasistatic, rectified components near zero frequency (Rect.), natural oscillations (Nat. Osc.), and sum-frequency generation (SFG). 
These features are captured by the mode-resolved electro-chiral susceptibility, $|\chi_{c,xy}^{(2)}(\omega,\omega_0)|$, shown in (c,g). 
(d,h) Structural distortions induced by the $A_2$ modes upon rectification, viewed along the $z$-axis of the crystal. Li ions are shown in blue, O ions in red, and B ions in green. The $A_2$(3.5) mode corresponds to a breathing mode of the borate framework, whereas the $A_2$(44.3) mode involves stretching and bending of the B-O bonds.
}
\end{figure*}


\subsection*{Phonon-induced geometric chirality in lithium triborate}

We now evaluate the theory developed in the previous sections for the example of LiB$_3$O$_5$, an ionic insulator that is well-suited for phonon driving. LiB$_3$O$_5$ crystallizes in the $Pna2_1$ space group (point group $C_{2v}$) and the 36-atom unit cell hosts 108 phonon modes with irreducible representations $A_1$, $A_2$, $B_1$, and $B_2$. The $A_2$ modes in the system break all mirror symmetries and reduce the space group of the material to $P2_1$ (point group $C_2$), which is one of the Sohncke groups. Nonlinear phononic rectification along the eigenvectors of an $A_2$ mode therefore creates a chiral crystal geometry. According to our analysis in Table~\ref{tab:phonons1c}, the $A_2$ modes couple to the IR-active $B_1$ and $B_2$ modes that can be driven coherently by an ultrashort laser pulse. We perform density functional theory calculations to obtain the phonon eigenfrequencies, eigenvectors, and nonlinear couplings. (See Methods for the computational details.)

We investigate resonant driving of the high-frequency transverse optic (TO) $B_1$ mode at $\Omega_a/(2\pi)=44.5$~THz and $B_2$ mode at $\Omega_b/(2\pi)=44.0$~THz. We use a mid-IR pulse with a peak electric field of $E_0=15$~MV/cm, pulse duration of $\tau=100$~fs, and center frequency of $\omega_0/(2\pi)=44.3$~THz. The pulse is polarized $\phi=45^\circ$ in the $xy$-plane of the orthorhombic crystal to drive both IR-active modes simultaneously, which couple to all 27 $A_2$ modes in the system. The $B_1$ and $B_2$ modes also couple to the fully symmetric $A_1$ modes, which leave the crystal symmetry unchanged and are therefore not considered here. We pick the illustrative examples of the lowest and highest frequency $A_2$ modes at 3.5 and 44.3~THz to demonstrate phonon-induced geometric chirality. The computed mode effective charges are $Z_{a,x}=-1.1$~$e$/$\sqrt{u}$ and $Z_{b,y}=-1.27$~$e$/$\sqrt{u}$, where $e$ is the elementary charge. The $A_2$ modes are not IR-active, $Z_{c,i}=0$. The computed coupling coefficients are $g=2$~meV/($\text{\AA}\sqrt{u}$)$^3$ for the $A_2$(3.5) mode and $g=56$~meV/($\text{\AA}\sqrt{u}$)$^3$ for the $A_2$(44.3) mode. We further assume phenomenological phonon linewidths of $\kappa_\alpha=0.1\Omega_\alpha/(2\pi)$ \cite{fechner:2016,juraschek2:2017}. 

We show the response of the system to excitation by the ultrashort mid-IR pulse as described by Eqs.~\eqref{eq:phononeom} for the $A_2$(3.5) mode in Figs.~\ref{fig:examples}(a-d) and for the $A_2$(44.3) mode in Figs.~\ref{fig:examples}(e-h). We plot the phonon amplitudes, $Q_c$, as well as the mean mode-resolved continuous chirality measures, $\langle S_c \rangle$, in Figs.~\ref{fig:examples}(a,e). Before the laser pulse hits, $t<0$, the crystal is in its achiral equilibrium characterized by space group $Pna2_1$. After the action of the pulse, $t>0$, the lattice is distorted impulsively along the eigenvectors of the $A_2$ modes in response to the coupling to the $B_1$ and $B_2$ modes. The phonon amplitudes show oscillations that are offset from their equilibrium position at $Q_c=0$. From the mean chirality measure $\langle S_c \rangle$, we see that the symmetry of the crystal structure is reduced to $P2_1$, a chiral state. $\langle S_c \rangle$ decays quadratically with the lifetime of the IR-active phonon modes and after $t=600$~fs, only the oscillatory parts of the $A_2$ modes remain, which means that on average, the crystal is back to its achiral $Pna2_1$ state. When the polarization of the pulse is rotated by 90$^\circ$ in the $xy$-plane of the crystal, the sign of the driving force $Q_a Q_b$ changes and accordingly so do those of $Q_c$ and $\langle S_c \rangle$. Because positive and negative displacements of the geometric chiral phonons are related by mirror symmetry, each of the enantiomers can be created selectively. Both $A_2$ modes show rectification, arising from the average force of the IR-active modes, $\langle Q_a Q_b \rangle$, but they show very different oscillatory behavior.

To disentangle the different features of the phonon dynamics, we show the Fourier transforms of the phonon amplitudes in Figs.~\ref{fig:examples}(b,f). There is a  rectified component near zero frequency, arising from the difference frequency of the $B_1$ and $B_2$ modes at $(\Omega_a-\Omega_b)/(2\pi)=0.5$~THz, which corresponds to the quasistatic distortion. The $A_2$(3.5) mode shows a clear peak at its eigenfrequency, which can be explained due to its relative proximity to the difference-frequency component of the $B_1$ and $B_2$ modes, which acts as the driving force in Eq.~\eqref{eq:phononeom}. The $A_2$(44.3) mode in turn feels the force of the sum-frequency component of the $B_1$ and $B_2$ modes at $(\Omega_a+\Omega_b)/(2\pi)=88.5$~THz, which corresponds to sum-frequency generation. All three features are captured by the electro-chiral susceptibility,  $\chi_{c,xy}^{(2)}$, which we show in Figs.~\ref{fig:examples}(c,g) for resonant driving conditions, $\omega'=\omega_0$. The peaks near zero frequency correspond to rectification, the peaks at the eigenfrequencies of the $A_2$ modes correspond to their natural oscillations, and the peaks at the sum frequency of the $B_1$ and $B_2$ modes correspond to sum-frequency generation, respectively \cite{Kahana2023}.

Finally, we show the structural distortions corresponding to the two $A_2$ modes in Figs.~\ref{fig:examples}(d,h). The low-frequency $A_2$(3.5) mode corresponds to a breathing mode of the borate framework, whereas the high-frequency $A_2$(44.3) mode involves stretching and bending of B--O bonds. Both are transiently frozen in by the rectification mechanism. In an experiment, the total structural distortion would be a superposition of all 27 $A_2$ modes of the system, showing varying components of the features displayed by the $A_2$(3.5) and $A_2$(44.3) modes, and with the chirality measure described by the total electro-chiral susceptibility (see Methods). In all calculations, we have evaluated the stability of the crystal with respect to the Lindemann criterion, predicting possible melting beyond a mean-square amplitude that displaces the atoms by more than 10\%{} of the interatomic distance. 
The largest displacements along the coordinates of the IR-active phonons in our simulations is 1~\%, well within the stability criterion and experimental achievability.


\section*{Discussion}

We have shown that geometric chiral phonon modes break all improper rotation symmetries, reducing the symmetry of the crystal structure from an achiral to a chiral point group. These phonon modes can be unidirectionally displaced through nonlinear phononic rectification in response to the excitation by an ultrashort laser pulse, leading to a transient, phonon-induced chiral state. Reorienting the polarization of the laser pulse switches the direction of the displacement, connected by mirror symmetry, enabling enantioselective generation of chirality. In our example of LiB$_3$O$_5$, the rectified $A_2$ modes reduce the crystal symmetry from space group $Pna2_1$ (point group $C_{2v}$) to $P2_1$ ($C_2$) and switching between the two enantiomers can be achieved by rotating the polarization of the pulse by 90$^\circ$ in the $xy$-plane of the orthorhombic crystal. The mechanism is generally applicable to materials in all 10 noncentrosymmetric-achiral point groups listed in Table~\ref{tab:phonons1c}. 

Ultrafast pump-probe experiments have proven to be able to measure chirality in molecules \cite{Rouxel2022,Ayuso2022}. A possible experiment measuring phonon-induced geometric chirality in solids could involve time-resolved electron diffraction following a mid-IR pump, which has proven to be a powerful tool to resolve phonon dynamics in recent years \cite{Tauchert2022}. In LiB$_3$O$_5$, the reduction in symmetry would lead to the appearance of Bragg peaks in diffraction patterns at $0kl$: $k + l = 2n + 1$, $h0l$: $h = 2n + 1$, $h00$: $h = 2n + 1$, and $0k0$: $k = 2n + 1$, where $n$ is any integer \cite{nespolo2017international}. A second possible experiment could involve X-ray chiral dichroism \cite{Mitcov2020,Freixas2023}, in which the two enantiomers accessed by a reorientation of the pulse polarization could be distinguished. A third possible experiment involves the changes of gyrotropic optical properties that follow the creation of chirality. Changing the point group from $C_{2v}$ to $C_2$ in LiB$_3$O$_5$ introduces diagonal components to the gyrotropic tensor \cite{He2020_gyrotropic}, which can be measured in a terahertz pump-optical probe setup. Other promising candidates for this type of measurement involve compounds of the $C_{3v}$ point group (e.g. LiNbO$_3$), in which a rectification of the $A_2$ modes reduces the symmetry to $C_{3}$, introducing diagonal gyrotropic tensor components. Furthermore, it is possible to induce gyrotropy in nongyrotropic materials, for example by rectifying the $E$ modes in point group $T_d$ (nongyrotropic), which reduces the symmetry to $D_2$ (gyrotropic). 

Our prediction of phonon-induced geometric chirality and enantioselective switching introduces a new paradigm in the optical control of solids, by enabling the engineering of chiral optical properties on demand and potentially allowing to induce chiral electronic phases, including chiral topology and superconductivity. The mechanism of inducing geometric chirality in solids further constitutes a novel path to controlling magnetoelectric order (e.g. ferrotoroidal and ferroaxial) and chiral magnetism \cite{ding2021field,Cheong2022,He2023}.


\section*{Methods}

\small

\subsection*{Computational details}

We computed the phonon frequencies and eigenvectors at the Brillouin-zone center, as well as the Born effective charges using density functional perturbation theory (DFPT) as implemented in the \textit{ab-initio} software package \textsc{Abinit} \cite{gonze2020abinit}. We used the Perdew--Burke--Ernzerhof (PBE) exchange--correlation functional with the dispersion correction of Grimme \cite{grimme2010consistent}. For the plane-wave basis set, we used an energy cutoff of 38 Ha together with norm-conserving pseudopotentials obtained from the \textsc{Abinit} library, and a $6 \times 6 \times 6$ Monkhorst-Pack grid to sample the Brillouin zone. A structural relaxation was performed prior to the DFPT calculations to an internal pressure of $-10$~kPa, yielding lattice parameters $a = 8.37$ \AA, $b = 7.35$ \AA, and $c = 5.22$ \AA, consistent with experimental data \cite{pylneva1999growth}. We used the eigenvectors obtained with DFPT to compute the lattice energy of the system as a function of the displacements of the $B_1$, $B_2$, and $A_2$ modes on a $9 \times 9 \times 9$ grid covering $\pm 2$~\AA{}$\sqrt{u}$. We extracted the coupling constant $g$ from a 6th-order polynomial fit to this grid using the polyfitn toolbox in MATLAB. \newline{}

\subsection*{Solutions of the equations of motion} 

The equations of motion for the phonon amplitudes are given by
\begin{equation}
    \ddot{Q}_\alpha + \kappa_\alpha \dot{Q}_\alpha + \partial_{Q_\alpha}V  = 0, \nonumber
\end{equation}
where $\kappa_\alpha$ are the phonon linewidths, $\alpha\in\{a,b,c\}$, and $V=V_{ph}+V_{l\text{-}m}$ from the main text. In detail, they read
\begin{align}
\ddot{Q}_a + \kappa_a \dot{Q}_a + \Omega_a^2 Q_a & = -g Q_b Q_c + Z_{a,i}E_i, \nonumber\\
\ddot{Q}_b + \kappa_b \dot{Q}_b + \Omega_b^2 Q_b & = -g Q_a Q_c + Z_{b,i}E_i, \nonumber\\
\ddot{Q}_c + \kappa_c \dot{Q}_c + \Omega_c^2 Q_c & = -g Q_a Q_b. \nonumber
\end{align}
The solutions of these equations of motion in frequency space can be approximated to first and second order in the electric field for the IR-active and geometric chiral phonon modes, respectively \cite{Kahana2023}. They yield
\begin{align}
    Q_a(\omega) &= Z_{a,i}\frac{E_i(\omega)}{\Delta_a(\omega)}, \nonumber\\
    Q_b(\omega) &= Z_{b,i}\frac{E_i(\omega)}{\Delta_b(\omega)}, \nonumber\\
    Q_c(\omega) &=  -\frac{g}{\Delta_c(\omega)} \left({Q}_a(\omega) \circledast  {Q}_b(\omega)\right) \nonumber\\
     &= -\frac{gZ_{a,i}Z_{b,j}}{\Delta_c(\omega)}
     \left( \frac{E_i(\omega)}{\Delta_a(\omega)} \circledast \frac{E_j(\omega)}{\Delta_b(\omega)}\right), \nonumber
\end{align}
where $\circledast$ is the convolution operator and $\Delta_{\alpha}(\omega)=\Omega_{\alpha}^2 - \omega^2 + i\omega\kappa_{\alpha}$. \newline{}

\subsection*{Total electro-chiral susceptibility} 

If multiple geometric chiral phonon modes, labeled $c_l$, are involved in the nonlinear phonon dynamics, the continuous chirality measure,
\begin{equation}
s = \sqrt{\sum\limits_n ||\mathbf{r}_n-\mathbf{r}_{n,\mathrm{achiral}}||^2}, 
\end{equation}
contains a sum over them, $||\mathbf{r}_n-\mathbf{r}_{n,\mathrm{achiral}}||=  \left\vert\left\vert\sum_l Q_{c_l}\mathbf{q}_{c_l,n}\right\vert\right\vert/\sqrt{M_n}$. Normalizing by the eigenvectors of all modes involved, we can then write the total electro-chiral susceptibility as 
\begin{align}
\chi^{(2)}_{\mathrm{tot},ij}(\omega,\omega') = & \frac{1}{\text{\AA}\sqrt{u}}\frac{Z_{a,i}Z_{b,j}}{\Delta_a(\omega-\omega')\Delta_b(\omega')}\nonumber\\
& \times \sqrt{\frac{\sum\limits_n \frac{1}{M_n}||\sum\limits_l \frac{g_l}{\Delta_{c_l}(\omega)}\mathbf{q}_{c_l,n}||^2}{\sum\limits_n \frac{1}{M_n}||\sum\limits_l \mathbf{q}_{c_l,n}||^2}}. \nonumber
\end{align}
This expression reduces to the mode-resolved electro-chiral susceptibility in Eq.~\eqref{eq:electrochiralsusceptibility} of the main text when only one mode $c_l\equiv c$ is considered.


\begin{acknowledgments}
We are grateful to N. Spaldin, S. Bhowal, D. Bustamante Lopez, M. Horn von Hoegen, K. Ollefs, U. Staub, H. Kusunose, and A. Eschenlohr for useful discussions. We acknowledge support from Tel Aviv University and ETH Zurich. Computational resources were provided by the Swiss National Supercomputing Center (CSCS) under project ID s1128.
\end{acknowledgments}



%

\clearpage{}




\onecolumngrid

\setcounter{page}{1}

\begin{center}
\textbf{\large Supporting Information: Phonon-induced geometric chirality}\\[0.4cm]
Carl\ P.\ Romao,$^{1}$ and Dominik\ M.\ Juraschek$^{2}$\\[0.15cm]

$^1${\itshape{\small Department of Materials, ETH Zurich, CH-8093 Zurich, Switzerland}}\\
$^2${\itshape{\small School of Physics and Astronomy, Tel Aviv University, Tel Aviv 6997801, Israel}}\\
\end{center}

\setcounter{equation}{0}
\setcounter{figure}{0}
\setcounter{table}{0}
\setcounter{section}{0}
\makeatletter
\renewcommand{\theequation}{S\arabic{equation}}
\renewcommand{\thefigure}{S\arabic{figure}}
\renewcommand{\thetable}{S\arabic{table}}


\begin{table*}[h]
\centering
\begin{tabular}{lll} 
\hline\hline
Point group~~ & Space groups & Space group \# \\
\hline\hline
$C_1$(1) & $P_1$ & 1 \\
$C_2$(2) &  $P2$, $P2_1$, $C2$ &  3–5 \\
$D_2$(222) &  $P 222$, $P 222_1$, $P 2_1 2_1 2$, $P 2_1 2_1 2_1$, $C 222_1$, $C 222$, $F 222$, $I 222$, $I 2_1 2_1 2_1$ & 16–24 \\
$C_4$(4) &  $P4$, ${P 4_1}^*$, $P 4_2$, ${P 4_3}^*$, $I 4$, $I 4_1$ & 75–80 \\
$D_4$(422) &  $P 422$, $P 42_1 2$, ${P 4_1 22}^*$, ${P 4_1 2_1 2}^\dagger$, $P 4_2 22$, $P 4_2 2_1 2$, ${P 4_3 22}^*$, ${P 4_3 2_1 2}^\dagger$, $I 422$, $I 4_1 22$ & 89–98 \\
$C_3$(3) & $P 3$, ${P 3_1}^*$, ${P 3_2}^*$, $R 3$ & 143–146 \\
$D_3$(32) &  $P 312$, $P 321$, ${P 3_1 12}^*$, ${P 3_1 21}^\dagger$, ${P 3_2 12}^*$, ${P 3_2 21}^\dagger$, $R 32$ & 149–155 \\
$C_6$(6) &  $P 6$, ${P 6_1}^*$, ${P 6_5^*}$, ${P 6_2}^\dagger$, ${P 6_4}^\dagger$, $P 6_3$ &  168–173 \\
$D_6$(622) &  $P 622$, ${P 6_1 22}^*$, ${P 6_5 22}^*$, ${P 6_2 22}^\dagger$, ${P 6_4 22}^\dagger$, $P 6_3 22$ &  177–182 \\
$T$(23) & $P 23$, $F 23$, $I 23$, $P 2_1 3$, $I 2_1 3$ &  195–199 \\
$O$(432) &  $P 4_3 2$, $P 4_2 32$, $F 432$, $F 4_1 32$, $I 432$, ${P 4_3 32}^*$, ${P 4_1 32}^*$, $I 4_1 32$ & 207–214 \\
\multicolumn{3}{l}{{}$^{*\dagger}$Enantiomorphic space-group pairs} \\
\hline\hline
\end{tabular}
\caption{\textbf{Chiral point groups and Sohncke space groups.} The 11 chiral point groups host the 65 Sohncke groups that do not contain an improper rotation axis (including inversion, mirrors, and rotoreflections) and therefore describe chiral crystals. Of the Sohncke groups, 22 are enantiomorphic, meaning that a crystal structure and its enantiomorphic pair, which are interconverted by an odd number of mirror-symmetry operations, can be differentiated by the symmetry operations of their space group. For example, left-handed quartz crystallizes in space group $P3_221$ (no. 154) and right-handed quartz crystallizes in space group $P3_121$ (no. 152). In the other 43 Sohncke groups, the chirality of the crystal arises from the chirality of the asymmetric unit rather than that of the symmetry operations. Crystals in these space groups are nonhanded chiral, which means that pairs of enantiomers cannot unambiguously be assigned left- and right-handedness. This is also known as the Ruch shoe-potato problem \cite{king2003chirality}. Table adapted from Ref.~\cite{Fecher2022}. 
}
\label{tab:sohncke}
\end{table*}


\begin{table*}[h]
\centering
\begin{tabular}{lllllll} 
\hline\hline
Point group~ & $\mathcal{I}$~ & $\mathcal{M}$~ & Chiral mode~ &  Subgroup~ & Nonlinear phonon coupling & IR-active modes  \\
\hline\hline
Triclinic & & & &  & & \\
\hline
$C_1$(1) & \xmark & \xmark &  &   &    &   \\
$C_i$($\bar{1}$) & \cmark & \xmark &  $A_u$ & $C_1$(1) & $A_u A_u A_g$ & $A_u(x,y,z)$ \\
\hline
Monoclinic & & & & & \\
\hline
$C_2$(2) & \xmark & \xmark &  &   &    &   \\
$C_s$(m) & \xmark & \cmark & $A''$ & $C_1$(1) & $A'' A'' A'$ & $A'(x,y)$, $A''(z)$\\ 
$C_{2h}$(2/m) & \cmark & \cmark & $A_u$ & $C_2$(2) & $A_u A_u A_g$, $A_u B_u B_g$ & $B_u(x,y)$, $A_u(z)$\\
\hline
Orthorhombic & & &  & & & \\
\hline
$D_2$(222) & \xmark & \xmark &  &   &    &   \\
$C_{2v}$(mm2) & \xmark & \cmark & $A_2$ & $C_2$(2) & $A_2 A_2 A_1$, $A_2 B_1 B_2$ &  $B_1(x)$, $B_2(y)$, $A_1(z)$ \\
$D_{2h}$(mmm) & \cmark & \cmark & $A_u$ & $D_2$(222) & $A_u A_u A_g$, $A_u B_{\alpha u} B_{\alpha g}$ ($\alpha=1,2,3$) & $B_{3u}(x)$, $B_{2u}(y)$, $B_{1u}(z)$ \\
\hline
Tetragonal & & &  & & & \\
\hline
$C_4$(4) & \xmark & \xmark &  &   &    &   \\
$S_4$($\bar{4}$) & \xmark & \xmark & $B$ &  $C_2$(2)  & $BBA$, $B{}^{i}E{}^{i}E$  ($i=1,2$) &  ${}^{i}E(x,y)$, $B(z)$  \\
 & & & ${}^{i}E$ &  $C_1$(1)  &  ${}^{i}E{}^{j}EA$, ${}^{i}E{}^{i}EB$ ($i,j=1,2$; $i\neq j$) &    \\
$C_{4h}$(4/m) & \cmark & \cmark & $A_u$  & $C_4$(4) & $A_u A_u A_g$, $A_u B_u B_g$, $A_u {}^{i}E_u {}^{j}E_g $  & ${}^{i}E_u(x,y)$, $A_u(z)$ \\
 & & & & &  ($i,j=1,2$) & \\
& & & $B_u$ & $S_4$($\bar{4}$)$^*$ & $B_u B_u A_g$, $B_u A_u B_g$, $B_u {}^{i}E_u {}^{j}E_g$ &  \\
& & & & & ($i,j=1,2$; $i\neq j$)  & \\
$D_4$(422) & \xmark & \xmark &  &   &    &   \\
$C_{4v}$(4mm) & \xmark & \cmark & $A_2$ & $C_4$(4) & $A_2 A_2 A_1$, $A_2 B_1 B_2$, $A_2 E E$ & $E(x,y)$, $A_1(z)$ \\
 & & & $E$  & $C_1$(1), $C_s$(m)$^*$ & $E E A_\alpha$, $E E B_\alpha$ ($\alpha = 1,2$) & \\
$D_{2d}$($\bar{4}$2m) & \xmark & \cmark & $A_2$ & $S_4$($\bar{4}$)$^*$  & $A_2 A_2 A_1$, $A_2 B_1 B_2$, $A_2 E E$ & $E(x,y)$, $B_2(z)$ \\
& & & $B_1$ & $D_2$(222) & $B_1 B_\alpha A_\alpha$, $B_1 E E$ &  \\
 & & & $E$ & $C_1$(1), $C_2$(2), $C_s$(m)$^*$ & $E E A_\alpha$, $E E B_\alpha$ ($\alpha = 1,2$) & \\
$D_{4h}$(4/mmm) & \cmark & \cmark & $A_{1u}$ &  $D_4$(422) &  $A_{1u} A_{\alpha u} A_{\alpha g}$, $A_{1u} B_{\alpha u} B_{\alpha g}$, $A_{1u} E_u E_g$  & $E_u(x,y)$, $A_{2u}(z)$\\
& & & & & ($\alpha = 1,2$) & \\
\hline
Trigonal & & & &  & \\
\hline
$C_3$(3) & \xmark & \xmark &  &   &    &   \\
$C_{3i}$($\bar{3}$) & \cmark & \xmark & $A_u$ & $C_3$(3) & $A_u A_u A_g$, $A_u {}^{i}E_u {}^{i}E_g$ ($i=1,2$) & ${}^{i}E_u(x,y)$, $A_u$\\
& & & ${}^{i}E_u$ & $C_1$(1) & ${}^{i}E_u {}^{i}E_u A_g$, ${}^{i}E_u {}^{j}E_u A_g$, ${}^{i}E_u {}^{i}E_g A_u$,  &  \\
& & & & & ${}^{i}E_u {}^{j}E_g A_u$, ${}^{i}E_u {}^{j}E_u {}^{i}E_g$, ${}^{i}E_u {}^{i}E_u {}^{j}E_g$ & \\
& & & & & ($i,j=1,2$; $i\neq j$) & \\
$D_3$(32) & \xmark & \xmark &  &    &   &   \\
$C_{3v}$(3m) & \xmark & \cmark & $A_2$ & $C_3$(3) & $A_2 A_2 A_1$, $A_2 E E$ & $E(x,y)$, $A_1(z)$ \\ 
 & & & $E$ &  $C_1$(1), $C_s$(m)$^*$ & $E E A_\alpha$, $E E E$ ($\alpha = 1,2$) & \\
$D_{3d}$($\bar{3}$m) & \cmark & \cmark & $A_{1u}$ & $D_3$(32) &  $A_{1u} A_{\alpha u} A_{\alpha g}$, $A_{1u} E_u E_g$ ($\alpha = 1,2$) & $E_u(x,y)$, $A_{2u}(z)$\\
& & & $E_u$ & $C_1$(1), $C_2$(2), $C_s$(m)$^*$ & $E_u E_u A_{\alpha g}$, $E_u E_g A_{\alpha u}$, $E_u E_u E_g$  &  \\
& & & & & ($\alpha = 1,2$)  & \\
\multicolumn{7}{l}{{}$^*$Subgroup is not chiral} \\
\hline\hline
\end{tabular}
\caption{\textbf{Geometric chiral phonons by point group I.} We show the point groups of the triclinic, monoclinic, orthorhombic, tetragonal, and trigonal crystal systems. We mark the presence (\cmark) or absence (\xmark) of inversion and mirror symmetries, the irreducible representations of the geometric chiral phonons, the subgroups they lead to upon displacement, the possible three-phonon couplings relevant to nonlinear phononic rectification, and the irreducible representations of the IR-active phonon modes. For doubly and triply degenerate modes, the induced subgroup depends on the chosen basis of eigenvectors, which may also lead to achiral subgroups. Point group $S_4$($\bar{4}$) has no inversion or mirror symmetries, but still contains improper rotations, which makes it achiral.
}
\label{tab:phonons1a}
\end{table*}


\begin{table*}[h]
\centering
\begin{tabular}{lllllll} 
\hline\hline
Point group~ & $\mathcal{I}$~ & $\mathcal{M}$~ & Chiral mode~ &  Subgroup~  &  Nonlinear phonon coupling & IR-active modes  \\
\hline\hline
Hexagonal & & &  & & \\
\hline
$C_6$(6) & \xmark & \xmark &  &   &    &   \\
$C_{3h}$($\bar{6}$) & \xmark & \cmark & $A''$ & $C_3$(3) & $A'' A'' A'$, $A'' {}^{i}E' {}^{j}E''$ ($i,j=1,2$) & ${}^{i}E'(x,y)$, $A''(z)$ \\
 &  &  & ${}^{i}E''$ & $C_1$(1) & ${}^{i}E'' {}^{i}E'' A'$, ${}^{i}E'' {}^{j}E'' A'$, ${}^{i}E'' {}^{i}E' A''$, ${}^{i}E'' {}^{j}E' A''$,  &  \\
  & & &  & & ${}^{i}E'' {}^{i}E'' {}^{j}E'$ ($i,j=1,2$; $i\neq j$) & \\
$C_{6h}$(6/m) & \cmark & \cmark & $A_u$ & $C_6$(6) & $A_u A_u A_g$, $A_u B_u B_g$, $A_u {}^{i}E_{\alpha u} {}^{i}E_{\alpha g}$  & ${}^{i}E_{1u}(x,y)$, $A_u(z)$ \\
 & & & & & ($i=1,2$; $\alpha = 1,2$) & \\
& & & ${}^{i}E_{2u}$ & $C_2$(2) & ${}^{i}E_{2u} {}^{j}E_{2u} A_g$, ${}^{i}E_{2u} {}^{j}E_{2g} A_u$, ${}^{i}E_{2u} {}^{j}E_{1u} B_g$,    & \\
& & &        &          & ${}^{i}E_{2u} {}^{j}E_{1g} B_u$, ${}^{i}E_{2u} {}^{i}E_{\alpha u} {}^{j}E_{\alpha g}$, ${}^{i}E_{2u} {}^{j}E_{\alpha u} {}^{i}E_{\alpha g}$  & \\
 & & & & & ($i,j=1,2$; $\alpha = 1,2$)  & \\
$D_6$(662) & \xmark & \xmark &  &    &   &   \\
$C_{6v}$(6mm) & \xmark & \cmark & $A_2$ & $C_6$(6) & $A_2 A_2 A_1$, $A_2 B_1 B_2$, $A_2 E_\alpha E_\alpha$ ($\alpha=1,2$) & $E_1(x,y)$, $A_1(z)$\\
& & & $E_1$ & $C_1$(1), $C_s$(m)$^*$ & $E_1 E_1 A_\alpha$, $E_1 E_2 B_\alpha$, $E_1 E_1 E_2$ ($\alpha=1,2$) & \\
& & & $E_2$ & $C_2$(2), $C_{2v}$(mm2)$^*$ & $E_2 E_2 A_\alpha$, $E_2 E_1 B_\alpha$, $E_2 E_\alpha E_\alpha$ ($\alpha=1,2$) & \\
$D_{3h}$($\bar{6}$m2) & \xmark & \cmark & $A_1''$ & $D_3$(32) & $A_1'' A_\alpha'' A_\alpha'$, $A_1'' E' E''$ ($\alpha=1,2$) & $E'(x,y)$, $A_2''(z)$ \\
& & & $E''$ & $C_1$(1), $C_2$(2), $C_s$(m)$^*$ &  $E'' E'' A_\alpha'$, $E'' E' A_\alpha''$, $E'' E'' E'$ ($\alpha=1,2$) & \\
$D_{6h}$(6/mmm) & \cmark & \cmark & $A_{1u}$ & $D_6$(662) & $A_{1u} A_{\alpha u} A_{\alpha g}$, $A_{1u} B_{\alpha u} B_{\alpha g}$, $A_{1u} E_{\alpha u} E_{\alpha g}$  & $E_{1u}(x,y)$, $A_{2u}(z)$\\
& & & & &  ($\alpha = 1,2$) &  \\
& & & $E_{2u}$ & $C_2$(2), $D_2$(222),  & $E_{2u} E_{2u} A_{\alpha g}$, $E_{2u} E_{2g} A_{\alpha u}$, $E_{2u} E_{1u} B_{\alpha g}$,  & \\
& & & & $C_{2v}$(mm2)$^*$ &  $E_{2u} E_{1g} B_{\alpha u}$, $E_{2u} E_{\alpha u} E_{\alpha g}$ ($\alpha = 1,2$) & \\
\hline
Cubic & & & &  & & \\
\hline
$T$(23) & \xmark & \xmark &  &  &     &   \\
$T_{h}$(m$\bar{3}$) & \cmark & \cmark & $A_u$ & $T$(23) & $A_u A_u A_g$, $A_u {}^{i}E_u {}^{i}E_g$, $A_u T_u T_g$ ($i=1,2$)  & $T_u(x,y,z)$ \\
& & & ${}^{i}E_u$ & $D_2$(222) & ${}^{i}E_u {}^{j}E_u A_g$, ${}^{i}E_u {}^{j}E_g A_u$, ${}^{i}E_u {}^{i}E_u {}^{j}E_g$,   & \\
 & & & &  & ${}^{i}E_u {}^{j}E_u {}^{i}E_g$, ${}^{i}E_u T_u T_g$ ($i,j=1,2$) & \\
$O$(432) & \xmark & \xmark &  &    &   &   \\
$T_d$($\bar{4}$3m) & \xmark & \cmark & $A_2$ & $T$(23) & $A_2 A_2 A_1$, $A_2 E E$, $A_2 T_1 T_2$ & $T_2(x,y,z)$ \\
& & & $E$ & $D_2$(222), $D_{2d}$($\bar{4}$2m)$^*$ & $E E A_\alpha$, $E E E$, $E T_\alpha T_\beta$ ($\alpha,\beta = 1,2$) & \\
& & & $T_1$ & $P_1$(1), $C_s$(m),$^*$ $S_4$($\bar{4}$),$^*$  & $T_1 T_1 A_1$, $T_1 T_2 A_2$, $T_1 T_1 E$, $T_1 T_\alpha T_\beta$  & \\
& & & & $C_3$(3) & ($\alpha,\beta = 1,2$) & \\
$O_{h}$(m$\bar{3}$m) & \cmark & \cmark & $A_{1u}$ & $O$(432) & $A_{1u} A_{\alpha u} A_{\alpha g}$, $A_{1u} E_u E_g$, $A_{1u} T_{\alpha u} T_{\alpha g}$  & $T_{1u}(x,y,z)$ \\
 & & & & & ($\alpha=1,2$) & \\ 
& & & $E_u$ & $D_2$(222), $D_4$(422),   & $E_u E_u A_{\alpha g}$, $E_u E_g A_{\alpha u}$, $E_u E_u E_g$, $E_u T_{\alpha u} T_{\beta g}$  & \\
& & & & $D_{2d}$($\bar{4}$2m)$^*$ & ($\alpha,\beta=1,2$)  & \\
\multicolumn{7}{l}{{}$^*$Subgroup is not chiral} \\
\hline\hline
\end{tabular}
\caption{\textbf{Geometric chiral phonons by point group II.} We show the point groups of the hexagonal and cubic crystal systems. We mark the presence (\cmark) or absence (\xmark) of inversion and mirror symmetries, the irreducible representations of the geometric chiral phonons, the subgroups they lead to upon displacement, the possible three-phonon couplings relevant to nonlinear phononic rectification, and the irreducible representations of the IR-active phonon modes. For doubly and triply degenerate modes, the induced subgroup depends on the chosen basis of eigenvectors, which may also lead to achiral subgroups.
}
\label{tab:phonons1b}
\end{table*}

\end{document}